\documentclass[10pt]{article}
\usepackage{graphicx}
\usepackage{amsmath}
\usepackage{amssymb}
\usepackage{caption2}
\begin{document}
\bibliographystyle{prsty}
\begin{center}
{\large {\bf \sc{ Analysis of the radiative decays among the charmonium states  }}} \\[2mm]
Zhi-Gang Wang  \footnote{E-mail:wangzgyiti@yahoo.com.cn. } \\
  Department of Physics, North China Electric Power University, Baoding 071003, P. R.
  China
\end{center}

\begin{abstract}
In this article, we study the radiative decays among the charmonium
states with the heavy quark effective theory, and make predictions
for the ratios among the radiative decay widths of an special
multiplet to another multiplet. The predictions can be confronted
with the experimental data in the future and put additional
constraints in identifying  the $X$, $Y$, $Z$ charmonium-like
mesons.
\end{abstract}


\section{Introduction}

In recent years, the Babar, Belle, CLEO, D0, CDF and FOCUS
collaborations have discovered (or confirmed) a large number of
charmonium-like states
  and revitalized  the interest
in the spectroscopy of the charmonium states. Many possible
assignments for those states have been suggested, such as the
conventional charmonium states, the multiquark states (irrespective
of the molecule type and the diquark-antidiquark type), the hybrid
states, the baryonium states, the threshold effects, etc
\cite{Swanson06XYZ,Olsen-XYZ,ReviewKlempt,ReviewVoloshin,ReviewEichten,ReviewZhu,Recent-review}.
The main difficulties in identifying  those  new mesons as the
excited charmonium states in that the observed masses do not fit in
with  the predictions of the potential models based on the
confining  potential. In this article, we will focus on the
traditional charmonium scenario and study the radiative decays among
the charmonium states with the heavy quark effective theory
\cite{PRT1997,RevWise,RevNeubert}.

In 2003, the Belle collaboration observed the $X(3872)$ in the
$J/\psi\pi^{+}\pi^{-}$ invariant mass distribution with a
significance in excess of $10\,\sigma$ in the $B\to K
J/\psi\pi^{+}\pi^{-}$ decay \cite{Belle0309032}, and later the
$X(3872)$ was confirmed by the D0, CDF and Babar collaborations
\cite{ConfirmX3872CDF,ConfirmX3872D0,ConfirmX3872Babar}.
The   decay $X(3872)\to J/\psi \gamma$ observed by the Belle and Babar collaborations
\cite{Belle05X3872JPC,Babar06X3872JPC,Babar09X3872JPC} and the decay   $X(3872)\to \psi' \gamma$ observed by the  Babar collaboration
\cite{Babar09X3872JPC} favor the even charge conjunction assignment.
 The di-pion spectrum also indicates that the charge conjunction  is even
\cite{Belle0309032}, and the studies of the   Belle collaboration
\cite{Belle05X3872JPC}  and CDF collaboration
\cite{CDF06X3872JPC-1,CDF06X3872JPC-2} favor the spin-parity-charge-conjunction
$J^{PC}=1^{++}$ assignment for the $X(3872)$. However, the recent
analysis of the $B\to J/\psi \omega K$ decay  by the Babar
collaboration indicates that the $P$-wave orbital angular momentum
for the $J/\psi \omega$ system is more favored than the $S$-wave,
the $X(3872)$ maybe have  the spin-parity-charge-conjunction  $J^{PC}=2^{-+}$ instead of
the $J^{PC}=1^{++}$ \cite{delAmoSanchez2010X3872}. In
Ref.\cite{Jia2010}, Jia et al study the   radiative decays
$X(3872)\to J/\psi(\psi')+\gamma$ within several phenomenological
potential models, and observe that the $2^{-+}$ assignment for the
$X(3872)$ is highly problematic. In this article, we assume that the
$X(3872)$ is the conventional charmonium $\chi_{c1}(2{\rm P})$
\cite{X3872-th-1,X3872-th-2,X3872-th-3,QuiggX3872,LiChao2009}.

In 2005, the Belle collaboration observed the $Z(3930)$   in the
$D\bar{D}$ invariant mass distribution  near $3.93\,\rm{GeV}$ in the
$\gamma \gamma$ collision with the statistical significance of
$5.3\,\sigma$ \cite{Belle05Z3930}, the mass and width are
 $M_{Z(3930)}=(3929 \pm 5 \pm 2) \,\rm{MeV} $ and $\Gamma_{Z(3930)} = (29 \pm 10\pm
 2) \,\rm{MeV}$, respectively. The production rate and the angular distribution favor  the
 $\chi_{c2}(2{\rm P})$ assignment \cite{Belle05Z3930}.
 In the same year,  the  Belle collaboration observed the  $X(3940)$  in the
recoiling spectrum of the $J/\psi$ in the  process $e^+e^-\to J/\psi
+D^*\bar{D}$ \cite{Belle05X3940}. Later the Belle collaboration
 studied it with higher statistics and determined the mass and
 width $ M_{X(3940)}=\left(3942_{-6}^{+7}\pm6 \right)\,\rm{MeV}$ and
   $\Gamma_{X(3940)}=\left(37_{-15}^{+26}\pm8\right)
  \,\rm{MeV}$ \cite{Belle08X3940X4160}. Furthermore, they
   observed the $X(4160)$ in the $D^*\bar{D}^*$ invariant mass
   distribution in the process $e^+e^-\to J/\psi +D^*\bar{D}^*$ with a significance
of  $5.1\,\sigma$. The mass and width of the $X(4160)$ are
$M_{X(4160)}=\left(4156_{-20}^{+25}\pm15\right)\,\rm{MeV}$ and
$\Gamma_{X(4160)}=\left(139_{-61}^{+111}\pm21 \right) \, \rm{MeV}$,
respectively.  The observation of the dominant decay mode of the
$X(3940)$ being the $D^*\bar{D}$ and the lack of evidence for the
$D\bar{D}$ decay mode indicate that it is a good candidate for the
$\eta_c(3{\rm S})$ state
\cite{Belle05X3940,Belle08X3940X4160,RosnerX3940}, although the mass
is lower than the prediction of the potential models
\cite{Barnes2005}. The dominant decay mode of the $X(4160)$ is the
$D^*\bar{D}^*$ \cite{Belle08X3940X4160}, and the $D\bar{D}$ and
$D\bar{D}^*$ modes have not been observed, so the $X(4160)$ may be
the $\eta_c(4{\rm S})$ or $\chi_{c0}(3{\rm P})$ candidate
\cite{Chao08X4160}.

Also in 2005, the Babar collaboration observed the $Y(4260)$ in the
$J/\psi\pi^{+}\pi^{-}$ invariant mass distribution in the initial
state radiation (ISR) process $e^+e^- \to \gamma_{\rm ISR}
J/\psi\pi^{+}\pi^{-}$ \cite{BaBar05Y4260}. It was confirmed by the
CLEO collaboration \cite{CLEO06Y4260} and Belle collaboration
\cite{Belle07Y4008Y4260}, and in the Belle data there is also a
relative broad structure $Y(4008)$ with the mass
$M_{Y(4008)}=\left(4008\pm 40^{+114}_{-28}\right)\,\rm{MeV}$ and
width $\Gamma_{Y(4008)}=\left(226\pm 44\pm 87 \right) \, \rm{MeV}$,
respectively. In Ref.\cite{Llanes-EstradaY4260}, Llanes-Estrada
suggests that the $Y(4260)$ may be the $\psi(4{\rm S})$  charmonium
state and displaces  the $\psi(4415)$.

In 2006, the Babar collaboration  observed the structure $Y(4320)$
 in the initial
state radiation  process $e^+e^- \to \gamma_{\rm ISR} \psi'
\pi^{+}\pi^{-}$ with the mass $M_{Y(4320)}=\left(4324\pm24 \right)
\, \rm{MeV}$ and width
$\Gamma_{Y(4320)}=\left(172\pm33\right)\,\rm{MeV} $, respectively
\cite{BaBar07Y4325}, and later the Belle collaboration observed two
relative narrow resonant structures  $Y(4360)$ and $Y(4660)$ with
the masses $M_{Y(4360)}=\left(4361\pm9\pm9\right) \, \rm{MeV}$,
$M_{Y(4660)}=\left(4664\pm11\pm5\right) \, \rm{MeV}$ and
 the widths $\Gamma_{Y(4360)}=\left(74\pm15\pm10 \right) \, \rm{MeV}$,
 $\Gamma_{Y(4660)}=\left(48\pm15\pm3 \right) \,\rm{MeV}$, respectively \cite{Belle07Y4350Y4660}.
 In 2008, the Belle collaboration observed  the  $Y(4630)$  as a
threshold enhancement in the $\Lambda_c^+\Lambda_c^-$ invariant mass
distribution in the initial state radiation process  $e^+e^- \to
\gamma_{\rm ISR} \Lambda_c^+\Lambda_c^-$ \cite{Belle08Y4630}. The
mass and width are $M_{Y(4630)}=\left(4634_{-7-8}^{+8+5}\right)\,
\rm{MeV}$ and $\Gamma_{Y(4630)}=\left(92_{-24-21}^{+40+10}\right)\,
\rm{MeV}$, respectively, which are roughly in agreement with that of
the $Y(4660)$.  In Ref.\cite{LiChao2009}, Li and Chao calculate the
mass spectrum of the charmonium states based on the screened
potential, and observe that the $Z(3930)$ agrees with the
$\chi_{c2}(2{\rm P})$, the $\psi(4415)$ is compatible with the
$\psi(5{\rm S})$ rather than the $\psi(4{\rm S})$, the $Y(4260)$,
$Y(4360)$ and $Y(4660)$ may be  the $\psi(4{\rm S})$, $\psi(3{\rm
D})$ and $\psi(6{\rm S})$ respectively, and the $X(3940)$ and
$X(4160)$ may be  the $\eta_c(3{\rm S})$ and $\chi_{c0}(3{\rm P})$
respectively.  In Ref.\cite{Ding08Y4350Y4660}, Ding et al identify
the  $Y(4360)$ and $Y(4660)$ as the  $\rm 3 ^3{\rm D}_1$ and $ 5
^3{\rm S}_1$ charmonium  states respectively (Badalian et al share
the same opinion \cite{BadalianY4350Y4660}), and evaluate the
$e^{+}e^{-}$ leptonic widths, $E1$ transitions, $M1$ transitions and
the open flavor strong decays.

In 2009,  the CDF collaboration observed a narrow structure
$Y(4140)$ near the $J/\psi\phi$ threshold with a statistical
significance in excess of $3.8\,\sigma$  in the exclusive $B\to
J/\psi\phi K$ decays produced in $\bar{p} p $ collisions
 \cite{CDF2009Y4140}. The mass and width are
$M_{Y(4140)}=\left(4143.0\pm2.9 \pm1.2\right)\,\rm{MeV}$ and
$\Gamma_{Y(4140)}=\left(11.7^{+8.3}_{-5.0}\pm3.7\right)\,\rm{MeV}$,
respectively \cite{CDF2009Y4140}. The $Y(4140)$ is very similar to
the charmonium-like state $Y(3940)$,  which was observed by both the
Belle and Babar collaborations near the $J/\psi\omega$ threshold in
the exclusive $B\to J/\psi\omega K$ decays
\cite{BelleY3940,BabarY3940}. The mass and width are
$M_{Y(3940)}=\left(3943\pm11 \pm 13\right) \,\rm{MeV}$ and
$\Gamma_{Y(3940)} =\left( 87 \pm 22\pm 26\right)\,\rm{ MeV}$
respectively from the Belle collaboration \cite{BelleY3940}
 and $M_{Y(3940)}=\left(3914.6 ^{+3.8}_{-3.4}
\pm{2.0} \right)\,\rm{MeV}$ and
$\Gamma_{Y(3940)}=\left(34^{+12}_{-8} \pm{5} \right)\,\rm{MeV}$
respectively from the Babar collaboration \cite{BabarY3940}.
In 2009, the Belle collaboration reported the observation of a significant enhancement  with the mass $\left(3915 \pm 3 \pm 2 \right)\,\rm{MeV}$ and total width $ \left(17 \pm 10\pm 3\right)\,\rm{MeV} $ respectively in the process $\gamma \gamma \to \omega J/\psi$ \cite{Belle2010Y3940}, these values are
consistent with that of the $Y(3940)$.   The updated values of the  mass $\left(3919.1^{+3.8}_{-3.5} \pm 2.0\right)\,\rm{MeV}$  and total width $\left(31^{+10}_{-8}\pm 5\right)\,\rm{MeV}$ from the  Babar collaboration are also consistent with the old ones \cite{PRD2010Babar}.
The Belle collaboration measured the process $\gamma \gamma \to \phi
J/\psi$ for the $ J/\psi\phi$ invariant mass distributions between
the threshold and $5\,\rm{GeV}$, and observed a narrow peak
$X(4350)$ with a significance of $3.2\,\sigma$ \cite{BelleX4350}.
The mass and width
 are $M_{X(4350)}=\left(4350.6^{+4.6}_{-5.1}\pm 0.7\right)\,\rm{MeV}$ and
$\Gamma_{X(4350)}=\left(13.3^{+17.9}_{-9.1}\pm
4.1\right)\,\rm{MeV}$, respectively; and  no signal for the
$Y(4140)\to  J/\psi\phi$ structure was observed \cite{BelleX4350}.
It is difficult to identify the $Y(3940)$, $Y(4140)$ and $X(4350)$
as the conventional charmonium states \cite{WangX4350}.

In Table 1, we list the experimental values of the charmonium states
with the possible identifications compared with the theoretical
predictions \cite{LiChao2009,Barnes2005,PDG}. We do not mean that
 such assignments are correct and exclude other possibilities, and
just take it for granted for the moment and study the radiative
decays among the charmonium states with the heavy quark effective
theory \cite{PRT1997,RevWise,RevNeubert}, which have been  applied
to
  identify the excited $D_s$ and $D$ mesons, such as the $D_s(3040)$, $D_s(2700)$,
$D_s(2860)$, $D(2550)$, $D(2600)$, $D(2750)$ and $D(2760)$
\cite{Colangelo1001,Colangelo0710,Colangelo0607,Colangelo0511,Wang1009}.

\begin{table}
\begin{center}
\begin{tabular}{|cc|c|c|c|c|}
\hline\hline
 \multicolumn{2}{|c|}{State} &Experiment \cite{PDG}    & SP \cite{LiChao2009}    & NRP \cite{Barnes2005} &   GI \cite{Barnes2005} \\
\hline
1S &  $J/\psi(1^3{\rm S}_1) $ &   3096.916                  & 3097               & 3090 & 3098 \\
   &  $\eta_c(1^1{\rm S}_0) $ &   2980.3                    & 2979               & 2982 & 2975 \\
\hline
2S &  $\psi'(2^3{\rm S}_1)  $ &   3686.09                   & 3673               & 3672 & 3676 \\
   &  $\eta_c'(2^1{\rm S}_0)$ &   3637                      & 3623               & 3630 & 3623 \\
\hline
3S &  $\psi(3^3{\rm S}_1)   $ &   4039 $[\psi(4040)]$       & 4022               & 4072 & 4100 \\
   &  $\eta_c(3^1{\rm S}_0) $ &{\bf?}\,3942 $[X(3940)]$     & 3991               & 4043 & 4064 \\
\hline
4S &  $\psi(4^3{\rm S}_1)   $ & {\bf?}\,4263 $[Y(4260)]$    & 4273               & 4406 & 4450 \\
   &  $\eta_c(4^1{\rm S}_0) $ &                             & 4250               & 4384 & 4425 \\
\hline
5S &  $\psi(5^3{\rm S}_1)   $ & {\bf?}\,4421 $[\psi(4415)]$ & 4463               &      &      \\
   &  $\eta_c(5^1{\rm S}_0) $ &                             & 4446               &      &      \\
\hline
6S &  $\psi(6^3{\rm S}_1)   $ &{\bf?}\,4664 $[Y(4660)]$     & 4608               &      &      \\
   &  $\eta_c(6^1{\rm S}_0) $ &                             & 4595               &      &      \\
\hline
1P &  $\chi_2(1^3{\rm P}_2) $ &  3556.20                    & 3554               & 3556 & 3550 \\
   &  $\chi_1(1^3{\rm P}_1 )$ &  3510.66                    & 3510               & 3505 & 3510 \\
   &  $\chi_0(1^3{\rm P}_0) $ &  3414.75                    & 3433               & 3424 & 3445 \\
   &  $h_c(1^1{\rm P}_1)    $ &  3525.42                    & 3519               & 3516 & 3517 \\
\hline
2P &  $\chi_2(2^3{\rm P}_2) $ &  3929 $[Z(3930)]$           & 3937               & 3972 & 3979 \\
   &  $\chi_1(2^3{\rm P}_1) $ &{\bf?}\,3871.56 $[X(3872)]$  & 3901               & 3925 & 3953 \\
   &  $\chi_0(2^3{\rm P}_0) $ &                             & 3842               & 3852 & 3916 \\
   &  $h_c(2^1{\rm P}_1) $    &                             & 3908               & 3934 & 3956 \\
\hline
3P &  $\chi_2(3^3{\rm P}_2) $ &                             & 4208               & 4317 & 4337 \\
   &  $\chi_1(3^3{\rm P}_1) $ &                             & 4178               & 4271 & 4317 \\
   &  $\chi_0(3^3{\rm P}_0) $ &{\bf?}\,4156 $ [X(4160)]$    & 4131               & 4202 & 4292 \\
   &  $h_c(3^1{\rm P}_1) $    &                             & 4184               & 4279 & 4318 \\
\hline
1D &  $\psi_3(1^3{\rm D}_3) $ &                             & 3799               & 3806 & 3849 \\
   &  $\psi_2(1^3{\rm D}_2) $ &                             & 3798               & 3800 & 3838 \\
   &  $\psi(1^3{\rm D}_1) $   &   3772.92 $[\psi(3770)]$    & 3787               & 3785 & 3819 \\
   &$\eta_{c2}(1^1{\rm D}_2)$ &                             & 3796               & 3799 & 3837 \\
\hline
2D &  $\psi_3(2^3{\rm D}_3) $ &                             & 4103               & 4167 & 4217 \\
   &  $\psi_2(2^3{\rm D}_2) $ &                             & 4100               & 4158 & 4208 \\
   &  $\psi(2^3{\rm D}_1) $   &  4153 $[\psi(4160)]$        & 4089               & 4142 & 4194 \\
   &$\eta_{c2}(2^1{\rm D}_2)$ &                             & 4099               & 4158 & 4208 \\
\hline
3D &  $\psi_3(3^3{\rm D}_3) $ &                             & 4331               &      &      \\
   &  $\psi_2(3^3{\rm D}_2) $ &                             & 4327               &      &      \\
   &  $\psi(3^3{\rm D}_1) $   &{\bf?}\,4361 $[Y(4360)]$     & 4317               &      &      \\
   &$\eta_{c2}(3^1{\rm D}_2)$ &                             & 4326               &      &      \\
\hline\hline
\end{tabular}
\end{center}
 \caption{Experimental and theoretical mass spectrum of the
charmonium states. The SP, NRP and GI denote the screened potential
model, the non-relativistic potential model and the Godfrey-Isgur
relativized potential model, respectively.}
 \end{table}

The article is arranged as follows:  we study  the   radiative
decays among the charmonium states with the heavy quark effective
theory in Sect.2; in Sect.3, we present the
 numerical results and discussions; and Sect.4 is reserved for our
conclusions.

\section{ The radiative decays with the heavy quark effective theory }

The  charmonium states can be classified according to the
 notation $n^{2s+1}L_{j}$, where the $n$ is the radial quantum number, the $L$ is
  the orbital angular momentum, the $s$ is  the spin, and the $j$ is the total angular
momentum. They have the parity and charge conjugation $P=(-1)^{L+1}$
and $C=(-1)^{L+s}$, respectively. The states have the same radial
quantum number $n$ and  orbital momentum $L$ can be expressed
by  the superfields $J$, $J^\mu$, $J^{\mu\nu}$, etc
\cite{GattoPLB93},
\begin{eqnarray}
J&=&\frac{1+{\rlap{v}/}}{2}\left\{\psi_{\mu}\gamma^\mu-\eta_c\gamma_5\right\}
\frac{1-{\rlap{v}/}}{2} \, , \nonumber \\
J^\mu&=&\frac{1+{\rlap{v}/}}{2}\left\{\chi_2^{\mu\nu}\gamma_\nu+\frac{1}{\sqrt{2}}\epsilon^{\mu\alpha\beta\lambda}v_\alpha
\gamma_{\beta}\chi^1_{\lambda}+\frac{1}{\sqrt{3}}\left(\gamma^\mu-v^\mu\right)\chi_0+h^\mu_c\gamma_5\right\}
\frac{1-{\rlap{v}/}}{2} \, , \nonumber \\
J^{\mu\nu}&=&\frac{1+{\rlap{v}/}}{2}\left\{\chi_3^{\mu\nu\alpha}\gamma_\alpha+\frac{1}{\sqrt{6}}
\left[\epsilon^{\mu\alpha\beta\lambda}v_\alpha
\gamma_{\beta}g^{\tau\nu}+\epsilon^{\nu\alpha\beta\lambda}v_\alpha
\gamma_{\beta}g^{\tau\mu}\right]\chi^2_{\tau\lambda}+\right.\nonumber\\
&&\left.\left[\sqrt{\frac{3}{20}}\left[\left(\gamma^\mu-v^\mu\right)g^{\nu\alpha}+\left(\gamma^\nu-v^\nu\right)g^{\mu\alpha}\right]
-\frac{1}{\sqrt{15}}\left(g^{\mu\nu}-v^{\mu}v^{\nu}\right)\gamma^\alpha\right]\chi^1_\alpha+\eta^{\mu\nu}_{c2}\gamma_5\right\}
\frac{1-{\rlap{v}/}}{2} \, , \nonumber\\
\end{eqnarray}
where the $v^{\mu}$ denotes the four-velocity associated to the
superfields, the fields  $\chi^3_{\mu\nu\alpha}$, $\chi^2_{\mu\nu}$,
$\chi^1_{\mu}$, $\chi^0$, $\eta_{c2}^{\mu\nu}$ have the total
angular momentum $j=3,2,1,0,2$ respectively, and belong to different
multiplets. The fields in a definite superfield have the same $n$, 
and form a multiplet. We multiply  the charmonium fields
$\chi^3_{\mu\nu\alpha}$, $\chi^2_{\mu\nu}$, $\chi^1_{\mu}$,
$\chi^0$, $\eta_{c2}^{\mu\nu}$, $\cdots$ with  a factor
$\sqrt{M_{\chi}}$, $\sqrt{M_{\eta}}$, $\cdots$, and they have
dimension of mass $\frac{3}{2}$.  The superfields
$J^{\mu_1\ldots\mu_L}$ have the following properties under the
parity, charge conjunction, heavy quark spin transformations,
\begin{eqnarray}
J^{\mu_1\ldots\mu_L}&\stackrel{P}{\longrightarrow}&\gamma^{0}J_{\mu_1\ldots\mu_L}\gamma^{0} \, ,  \nonumber \\
J^{\mu_1\ldots\mu_L}&\stackrel{C}{\longrightarrow}&(-1)^{L+1}C[J_{\mu_1\ldots\mu_L}]^{T}C \,,\nonumber \\
J^{\mu_1\ldots\mu_L}&\stackrel{S}{\longrightarrow}&SJ_{\mu_1\ldots\mu_L}S^{\prime\dagger}\,,\nonumber \\
v^{\mu}&\stackrel{P}{\longrightarrow}&v_{\mu}\, ,
\end{eqnarray}
where $S,S^{\prime}\in SU(2)$ heavy quark spin symmetry groups, and
$[S,{\rlap{v}/}]=[S^{\prime},{\rlap{v}/}]=0$.

The radiative transitions between the $ m$ and $ n$
charmonium states  can be described by the following Lagrangians,
\begin{eqnarray}
{\cal{L}}_{SS}&=&\sum_{m,n}\delta(m,n) \mathrm{Tr}\left[\bar{J}(m)\sigma_{\mu\nu}J(n)\right]F^{\mu\nu} \, , \nonumber\\
{\cal{L}}_{SP}&=&\sum_{m,n}\delta(m,n)\mathrm{Tr}\left[\bar{J}({m})J_{\mu}(n)+\bar{J}_\mu(n)
J(m)\right]V^{\mu} \, , \nonumber\\
{\cal{L}}_{PD}&=&\sum_{m,n}\delta(m,n)\mathrm{Tr}\left[\bar{J}_{\mu\nu}({m})J^{\nu}(n)+\bar{J}^\nu(n)
J_{\mu\nu}(m)\right]V^{\mu} \, ,
\end{eqnarray}
where $\bar{J}_{\mu_1\ldots\mu_L}=\gamma^0
J_{\mu_1\ldots\mu_L}^{\dag} \gamma^0$, $V^\mu=F^{\mu\nu}v_\nu$, the
$F^{\mu\nu}$ is the electromagnetic tensor, and the $\delta(m,n)$
are the coupling constants, which have different values in the
Lagrangians ${\cal{L}}_{SS}$, ${\cal{L}}_{SP}$, ${\cal{L}}_{PD}$, we
use the same notation for simplicity. The Lagrangians
${\cal{L}}_{SP}$ and ${\cal{L}}_{PD}$ preserve parity, charge
conjugation, gauge invariance and heavy quark spin symmetry, while
the Lagrangian ${\cal{L}}_{SS}$ violates the heavy quark symmetry.
There are two tensors
$\sigma_{\mu\nu}\widetilde{F}^{\mu\nu}(=\frac{1}{2}\epsilon^{\mu\nu\alpha\beta}\sigma_{\mu\nu}F_{\alpha\beta})$
and $\sigma_{\mu\nu}F^{\mu\nu}$ can be used to construct the
spin-breaking Lagrangian ${\cal{L}}_{SS}$.  They   both have the
terms $\vec{\sigma} \cdot \vec{E}$ and $\vec{\sigma} \cdot \vec{B}$,
and can turn into  each other with the interchange  $
\vec{E}\leftrightarrow \vec{B}$,  we choose the parity-conserving structure $\sigma_{\mu\nu}F^{\mu\nu}$ in
constructing the spin-breaking Lagrangian. We take the Lagrangian
${\cal{L}}_{SP}$ from Ref.\cite{PRT1997} and construct the
Lagrangians ${\cal{L}}_{SS}$ and ${\cal{L}}_{PD}$, and later  Dr. F.
De Fazio draws   my attention to Ref.\cite{Fazio2008}, where the
Lagrangian
 ${\cal{L}}_{PD}$ is introduced for the first time.

From the heavy quark effective Lagrangians  ${\cal L}_{SS}$, ${\cal
L}_{SP}$ and ${\cal L}_{PD}$, we can obtain the radiative decay
 widths $\Gamma$,
\begin{eqnarray}
\Gamma&=&\frac{1}{2j+1}\sum\frac{k_{\gamma}}{8\pi M_i^2 } |T|^2\, ,
\end{eqnarray}
where the $T$ denotes the scattering amplitude, the  $k_\gamma$ is
the momentum of the final states in the center of mass coordinate,
the $\sum$ denotes the sum of all the  polarization vectors,  the
$j$ is the total angular momentum of the initial state, and the $M_i$
is the mass of the initial state.   Cho and Wise study the
radiative decays of the heavy quarkonia using the multipole
expansion and heavy quark symmetry \cite{ChoWise}.

\section{Numerical Results}

We calculate the radiative decay widths $\Gamma$ using the FeynCalc
to carry out the sum of all the polarization  vectors. In
calculations, the masses of the charmonium states are taken as the
experimental values from the Review of Particle Physics \cite{PDG},
see Table 1; for the unobserved charmonium states, we take the
values from the screened potential model \cite{LiChao2009}, and
assume that  the charmonium states $\psi(5^3{\rm S}_1)$ and
$\eta_c(5^1{\rm S}_0)$ have degenerate masses. In this article, we
identify the   $\psi(4415)$ as the $\psi(5^3{\rm S}_1)$ charmonium
state, and expect that the $\eta_c(5^1{\rm
   S}_0)$ charmonium state  has  slightly smaller mass. The predictions of the screened
potential model are $M_{\psi(4{\rm S})}=4463\,\rm{MeV}$ and
$M_{\eta_c(4{\rm S})}=4446\,\rm{MeV}$, respectively, so we take the
approximation $M_{\psi(4{\rm S})}=M_{\eta_c(4{\rm
S})}=4421\,\rm{MeV}$.

 The numerical values of the radiative decay widths  are presented in Tables
 2-4, where we retain the unknown coupling constants $\delta(m,n)$
 among the multiplets of the radial quantum numbers $m$ and $n$. In general, we expect fitting
  the parameters $\delta(m,n)$  to the precise experimental data, however, in the
 present time the experimental data are far from enough. In Tables
 4-6, we present the ratios of the radiative decay widths among the
 charmonium states.

 The radiative decay widths listed in the Review of Particle Physics
  are \begin{eqnarray}
 \Gamma(\psi(2{\rm S})\to\eta_{c}(1{\rm{S}})\gamma)&=&1.0336\,\rm{KeV}\, ,\nonumber \\
 \Gamma(\psi(2{\rm S})\to\chi_{c0}(1{\rm{P}})\gamma)&=&29.2448\,\rm{KeV}\, ,\nonumber \\
 \Gamma(\psi(2{\rm S})\to\chi_{c1}(1{\rm{P}})\gamma)&=&27.968\,\rm{KeV}\, ,\nonumber \\
 \Gamma(\psi(2{\rm S})\to\chi_{c2}(1{\rm{P}})\gamma)&=&26.5696\,\rm{KeV}\, ,\nonumber \\
 \Gamma(\chi_{c0}(1{\rm{P}})\to J/\psi\gamma)&=&119.48\,\rm{KeV}\, ,\nonumber \\
 \Gamma(\chi_{c1}(1{\rm{P}})\to J/\psi\gamma)&=&295.84\,\rm{KeV}\, ,\nonumber \\
 \Gamma(\chi_{c2}(1{\rm{P}})\to J/\psi\gamma)&=&384.15\,\rm{KeV}\,,\nonumber \\
 \Gamma(\psi(3770)\to\chi_{c0}(1{\rm{P}})\gamma)&=&199.29\,\rm{KeV}\, ,\nonumber \\
 \Gamma(\psi(3770)\to\chi_{c1}(1{\rm{P}})\gamma)&=&79.17\,\rm{KeV}\, ,\nonumber \\
 \Gamma(\psi(3770)\to\chi_{c2}(1{\rm{P}})\gamma)&<&24.57\,\rm{KeV}\, ,\nonumber \\
 \Gamma(\psi(4040)\to\chi_{c1}(1{\rm{P}})\gamma)&<&0.88\,\rm{MeV}\, ,\nonumber \\
 \Gamma(\psi(4040)\to\chi_{c2}(1{\rm{P}})\gamma)&<&1.36\,\rm{MeV}\, ,
  \end{eqnarray}
 where we have neglected the uncertainties \cite{PDG}. From those radiative decay widths,
 we can obtain the following ratios,
\begin{eqnarray}
  \frac{\Gamma(\psi(2{\rm S})\to\chi_{c0}(1{\rm{P}})\gamma)}{\Gamma(\psi(2{\rm S})\to\chi_{c1}(1{\rm{P}})\gamma)}&=&1.046\,(1.151)\, ,\nonumber \\
  \frac{\Gamma(\psi(2{\rm S})\to\chi_{c0}(1{\rm{P}})\gamma)}{\Gamma(\psi(2{\rm S})\to\chi_{c2}(1{\rm{P}})\gamma)}&=&1.101\,(1.649)\, ,\nonumber \\
  \frac{\Gamma(\chi_{c0}(1{\rm{P}})\to J/\psi\gamma)}{\Gamma(\chi_{c2}(1{\rm{P}})\to J/\psi\gamma)}&=&0.311\,(0.363)\, ,\nonumber \\
  \frac{\Gamma(\chi_{c1}(1{\rm{P}})\to J/\psi\gamma)}{\Gamma(\chi_{c2}(1{\rm{P}})\to J/\psi\gamma)}&=&0.770\,(0.756)\, ,\nonumber \\
  \frac{\Gamma(\psi(3770)\to\chi_{c0}(1{\rm{P}})\gamma)}{\Gamma(\psi(3770)\to\chi_{c1}(1{\rm{P}})\gamma)}&=&2.517\,(3.167)\, ,\nonumber \\
  \frac{\Gamma(\psi(3770)\to\chi_{c0}(1{\rm{P}})\gamma)}{\Gamma(\psi(3770)\to\chi_{c2}(1{\rm{P}})\gamma)}&>&8.111\,(82.02)\, ,
  \end{eqnarray}
where the values in the bracket are the theoretical calculations
based on the heavy quark effective theory. The agreements  between
the experimental data and the theoretical calculations are rather
good, and the heavy quark effective theory works rather well.   The
ratios presented in Tables 4-6 can be confronted with the
experimental data in the future at the BESIII, KEK-B, RHIC,
$\rm{\bar{P}ANDA}$ and LHCb,  and put additional constraints  in
identifying  the $X$, $Y$, $Z$ charmonium-like mesons.

There is a relative $P$-wave between the  final-state charmonium and the photon, the radiative decay widths
$\Gamma \propto k_\gamma^3$, where  $k_\gamma=\frac{M_i^2-M_f^2}{2M_i}$, the $M_i$ and $M_f$ denote the
masses of the initial and final charmonium states respectively. The numerical values of the decay widths shown in Tables 2-4, where the effective coupling constants $\delta(m,n)$ have been  factorized  out,  reflect  the   corresponding processes are facilitated  or suppressed in the phase-space.
If the energy gap $M_i-M_f$ is small (or large),
small variations of the masses  $M_i$ and $M_f$ can  (or cannot) lead to remarkable changes for the decay width.  For example, we plot
the radiative decay widths  $\Gamma(X(3872)\to J/\psi \gamma)$ and $\Gamma(X(3872)\to \psi' \gamma)$ versus the mass $M_X$ in Fig.1.

  From the experimental data of the Babar collaboration  ${\rm {Br}}(B^+ \to X(3872) K^+) \times {\rm {Br}} (X(3872) \to J/\psi \gamma) = (2.8 \pm 0.8 \pm 0.2) \times 10^{-6}$ and ${\rm {Br}}(B^+ \to X(3872) K^+) \times {\rm {Br}}(X(3872) \to \psi' \gamma) = (9.9 \pm 2.9 \pm 0.6) \times 10^{-6}$ \cite{Babar09X3872JPC}, we can obtain the central value of the ratio $\frac{\Gamma(X(3872) \to J/\psi \gamma)}{\Gamma(X(3872) \to \psi' \gamma)}=0.283$, which means $\frac{\delta^2(2,1)}{\delta^2(2,2)}=0.00576$. The large hierarchy $\delta(2,1)\ll \delta(2,2)$ is compatible with the phenomenological expectation
    that the   $\chi_{c1}(\rm{2P})$ state potentially decays to the $\psi'\gamma$
rather  than to  the $J/\psi \gamma$,

The $E_1$ and $M_1$  transitions among the charmonium states are
usually calculated by the formula \cite{EM-1,EM-2,EM-3,E1-form},
\begin{eqnarray}
 \Gamma_{E1}\left({\rm n}^{2s+1}L_j
 \to{{\rm n}^\prime}^{2s'+1}{L'}_{j'}+\gamma\right)&=&\frac{4}{3}e_c^2\alpha
 E^3_\gamma \frac{E_f}{M_i}\delta_{ss'}C_{fi}\mid \langle {{\rm n}^\prime}^{2s'+1}{L'}_{j'}\mid r\mid {\rm n}^{2s+1}L_j
 \rangle\mid^2 \, , \nonumber \\
\Gamma_{M1}\left({\rm n}^{2s+1}L_j \to{{\rm
n}^\prime}^{2s'+1}{L'}_{j'}+\gamma\right)&=&\frac{4}{3}e_c^2\frac{\alpha}{m_c^2}
 E^3_\gamma \frac{E_f}{M_i}\frac{2j'+1}{2L+1}\delta_{LL'}\delta_{ss'\pm1}\mid \langle {{\rm n}^\prime}^{2s'+1}{L'}_{j'}\mid  {\rm n}^{2s+1}L_j
 \rangle\mid^2 \, ,\nonumber \\
\end{eqnarray}
where the $E_{\gamma}$ is the photon energy, the $E_f$ is the energy
of final state charmonium, the $ M_i$ is the mass of the initial
state charmonium, and the angular matrix factor $C_{fi}$ is
\begin{equation}
C_{fi}={\rm max}(L,  L')(2j'+1)\left\{
\begin{array}{ccc}
 L'&  J' &s\\
 J &  L  &1
\end{array}\right\}^2 \, .
\end{equation}
The values of the matrix elements  $\langle {{\rm
n}^\prime}^{2s'+1}{L'}_{j'}\mid r\mid {\rm n}^{2s+1}L_j
 \rangle$ and $\langle {{\rm
n}^\prime}^{2s'+1}{L'}_{j'}\mid {\rm n}^{2s+1}L_j
 \rangle$ depend on the details of the wave-functions which are evaluated
  using a special potential model, for example, the Cornell potential model, the logarithmic potential model,
  the power-law potential model, the QCD-motivated potential model \cite{4-potential}, the
  relativized Godfrey-Isgur model, the non-relativistic potential
model \cite{Barnes2005},  the screened potential model
\cite{LiChao2009}, the  relativistic quark model based on a
quasipotential approach in QCD \cite{Ebert2003,Ebert2005}, etc. All
predictions should be confronted with the experimental data.  In
this article, we intend to make estimations based on the heavy quark
effective theory, and prefer the ratios among the radiative decay
widths of an special multiplet to another multiplet, where the
unknown parameters $\delta(m,n)$ are canceled out with each other.

\begin{table}
\begin{center}
\begin{tabular}{|c|c|c|c|c|c| }\hline\hline
    & $\Gamma(\psi \to \chi_{c2}\gamma)$& $\Gamma(\psi \to \chi_{c1}\gamma)$ & $\Gamma(\psi \to \chi_{c0}\gamma)$ & $\Gamma(\eta_c \to h_c\gamma)$  \\ \hline
      $2{\rm S}\to1{\rm P}$ & 3.546    & 5.082     & 5.849      & 4.092  \\ \hline

      $3{\rm S}\to1{\rm P}$ & 146.4    & 112.1     & 57.15      & 174.9  \\ \hline
      $3{\rm S}\to2{\rm P}$ & 2.198    & 4.488     & 2.388      & 0.122  \\ \hline

      $4{\rm S}\to1{\rm P}$ & 405.8    & 287.4     & 126.3      & 768.9 \\ \hline
      $4{\rm S}\to2{\rm P}$ & 53.97    & 50.43     & 20.43      & 103.5   \\ \hline
      $4{\rm S}\to3{\rm P}$ & 0.285    & 0.620     & 0.407      & 0.880   \\ \hline

      $5{\rm S}\to1{\rm P}$ & 685.4    & 471.2     & 193.7      & 1323  \\ \hline
      $5{\rm S}\to2{\rm P}$ & 158.3    & 128.3     & 48.69      & 317.5    \\ \hline
      $5{\rm S}\to3{\rm P}$ & 15.13    & 13.26     & 5.647      & 36.96   \\ \hline

      $6{\rm S}\to1{\rm P}$ & 1283     & 858.8     & 328.0      & 2062   \\ \hline
      $6{\rm S}\to2{\rm P}$ & 466.5    & 341.7     & 122.7      & 695.3  \\ \hline
      $6{\rm S}\to3{\rm P}$ & 130.6    & 93.49     & 34.93      & 175.4    \\ \hline

      $1{\rm D}\to1{\rm P}$ & 0.258    & 6.691     & 21.19      &    \\ \hline
      $2{\rm D}\to1{\rm P}$ & 4.205    & 78.91     & 147.5      &    \\ \hline
      $2{\rm D}\to2{\rm P}$ & 0.288    & 8.305     & 14.63      &    \\ \hline

      $3{\rm D}\to1{\rm P}$ & 9.075    & 164.7     & 277.1      &    \\ \hline
      $3{\rm D}\to2{\rm P}$ & 1.821    & 38.97     & 60.39      &    \\ \hline
      $3{\rm D}\to3{\rm P}$ &0.097     & 2.438     & 4.506      &    \\ \hline
      \hline
\end{tabular}
\end{center}
\caption{ The radiative decay widths of the $S$-wave and $D$-wave
charmonium states to the $P$-wave charmonium states. The unit is
$10^{-4}\delta^2(m,n)$.  }
\end{table}

\begin{table}
\begin{center}
\begin{tabular}{|c|c|c|c|c|c| }\hline\hline
    & $\Gamma(\chi_{c2} \to\psi \gamma)$& $\Gamma( \chi_{c1}\to \psi\gamma)$ & $\Gamma(\chi_{c0} \to \psi\gamma)$ & $\Gamma(h_c \to \eta_c\gamma)$  \\ \hline
    $1{\rm P}\to1{\rm S}$ & 73.70      & 55.69     & 26.78      & 114.1  \\ \hline
    $2{\rm P}\to1{\rm S}$ & 350.4      & 294.9     & 260.5      & 442.2  \\ \hline
    $2{\rm P}\to2{\rm S}$ & 13.00      & 6.000     & 3.628      & 17.68   \\ \hline

    $3{\rm P}\to1{\rm S}$ & 720.8      & 686.1     & 624.0      & 827.2 \\ \hline
    $3{\rm P}\to2{\rm S}$ & 109.6      & 93.62     & 82.01      & 123.2  \\ \hline
    $3{\rm P}\to3{\rm S}$ & 4.628      & 2.621     & 1.583      & 12.97  \\ \hline   \hline
\end{tabular}
\end{center}
\caption{ The radiative decay widths of the $P$-wave charmonium
states to the $S$-wave charmonium states. The unit is
$10^{-4}\delta^2(m,n)$.  }
\end{table}

\begin{table}
\begin{center}
\begin{tabular}{|c|c|c|c|c| }\hline\hline
    & $\Gamma(\psi \to\eta_c \gamma)$& $\Gamma(\eta_c\to \psi\gamma)$ & $\frac{\Gamma(\eta_c\to \psi\gamma)}{\Gamma(\psi \to\eta_c \gamma)}$   \\ \hline
                   $2{\rm S}\to1{\rm S}$ & 0.089         & 0.136        & 1.519  \\ \hline

                   $3{\rm S}\to1{\rm S}$ & 0.244         & 0.430        & 1.762  \\ \hline
                   $3{\rm S}\to2{\rm S}$ & 0.021         & 0.018        & 0.848  \\ \hline

                   $4{\rm S}\to1{\rm S}$ & 0.384         & 0.919        & 2.392  \\ \hline
                   $4{\rm S}\to2{\rm S}$ & 0.071         & 0.161        & 2.281  \\ \hline
                   $4{\rm S}\to3{\rm S}$ & 0.012         & 0.011        & 0.911  \\ \hline

                   $5{\rm S}\to1{\rm S}$ & 0.502         & 1.273        & 2.536  \\ \hline
                   $5{\rm S}\to2{\rm S}$ & 0.127         & 0.325        & 2.550  \\ \hline
                   $5{\rm S}\to3{\rm S}$ & 0.035         & 0.057        & 1.614  \\ \hline
                   $5{\rm S}\to4{\rm S}$ & 0.002         & 0.005        & 2.384  \\ \hline

                   $6{\rm S}\to1{\rm S}$ & 0.712         & 1.692        & 2.375  \\ \hline
                   $6{\rm S}\to2{\rm S}$ & 0.253         & 0.561        & 2.221  \\ \hline
                   $6{\rm S}\to3{\rm S}$ & 0.106         & 0.160        & 1.505  \\ \hline
                   $6{\rm S}\to4{\rm S}$ & 0.024         & 0.039        & 1.616  \\ \hline
                   $6{\rm S}\to5{\rm S}$ & 0.005         & 0.006        & 1.143  \\ \hline \hline
\end{tabular}
\end{center}
\caption{ The  ratios of the radiative decay widths of the $S$-wave
charmonium states to the $S$-wave charmonium states. The unit of the
widths is $\delta^2(m,n)$. }
\end{table}

\begin{table}
\begin{center}
\begin{tabular}{|c|c|c|c|c|c| }\hline\hline
                                            &   $\Gamma(\psi \to \chi_{c1}\gamma)$ & $\Gamma(\psi \to \chi_{c0}\gamma)$ & $\Gamma(\eta_c \to h_c\gamma)$  \\ \hline
                      $2{\rm S}\to1{\rm P}$ &  1.433       & 1.649     & 1.154 \\ \hline
            $\widehat{2{\rm S}\to1{\rm P}}$ &  1.053       & 1.101     & \\ \hline

                      $3{\rm S}\to1{\rm P}$ &  0.766       & 0.390     & 1.195 \\ \hline
                      $3{\rm S}\to2{\rm P}$ &  2.042       & 1.087     & 0.056 \\ \hline

                      $4{\rm S}\to1{\rm P}$ &  0.708       & 0.311     & 1.895 \\ \hline
                      $4{\rm S}\to2{\rm P}$ &  0.934       & 0.379     & 1.918  \\ \hline
                      $4{\rm S}\to3{\rm P}$ &  2.176       & 1.428     & 3.090 \\ \hline

                      $5{\rm S}\to1{\rm P}$ &  0.688       & 0.283     & 1.931 \\ \hline
                      $5{\rm S}\to2{\rm P}$ &  0.810       & 0.308     & 2.006   \\ \hline
                      $5{\rm S}\to3{\rm P}$ &  0.876       & 0.373     & 2.443  \\ \hline

                      $6{\rm S}\to1{\rm P}$ &  0.669       & 0.256     & 1.607  \\ \hline
                      $6{\rm S}\to2{\rm P}$ &  0.732       & 0.263     & 1.490  \\ \hline
                      $6{\rm S}\to3{\rm P}$ &  0.716       & 0.267     & 1.343  \\ \hline

                      $1{\rm D}\to1{\rm P}$ &  25.91       & 82.02     &    \\ \hline
            $\widehat{1{\rm D}\to1{\rm P}}$ &  $>3.222$    & $>8.111$  &    \\ \hline
                      $2{\rm D}\to1{\rm P}$ &  18.77       & 35.08     &    \\ \hline
                      $2{\rm D}\to2{\rm P}$ &  28.85       & 50.81     &    \\ \hline

                      $3{\rm D}\to1{\rm P}$ &  18.15       & 30.54     &    \\ \hline
                      $3{\rm D}\to2{\rm P}$ &  21.40       & 33.17     &    \\ \hline
                      $3{\rm D}\to3{\rm P}$ &  25.27       & 46.71     &    \\ \hline
      \hline
\end{tabular}
\end{center}
\caption{ The ratios of the radiative decay widths of the $S$-wave
and $D$-wave charmonium states to the $P$-wave charmonium states.
There we normalize $\Gamma_{\psi \to \chi_{c2}\gamma}=1$, the
wide-hat denotes the experiential values. }
\end{table}

\begin{table}
\begin{center}
\begin{tabular}{|c|c|c|c|c|c| }\hline\hline
                                             & $\Gamma(\chi_{c1}\to \psi\gamma)$ & $\Gamma(\chi_{c0} \to \psi\gamma)$ & $\Gamma(h_c \to \eta_c\gamma)$  \\ \hline
                   $1{\rm P}\to1{\rm S}$     & 0.756     & 0.363    & 1.549 \\ \hline
         $\widehat{1{\rm P}\to1{\rm S}}$     & 0.770     & 0.311    &  \\ \hline
                   $2{\rm P}\to1{\rm S}$     & 0.842     & 0.744    & 1.262 \\ \hline
                   $2{\rm P}\to2{\rm S}$     & 0.462     & 0.279    & 1.360 \\ \hline

                   $3{\rm P}\to1{\rm S}$     & 0.952     & 0.866    & 1.148 \\ \hline
                   $3{\rm P}\to2{\rm S}$     & 0.854     & 0.748    & 1.124 \\ \hline
                   $3{\rm P}\to3{\rm S}$     & 0.566     & 0.342    & 2.803 \\ \hline   \hline
\end{tabular}
\end{center}
\caption{ The  ratios of the radiative decay widths of the $P$-wave
charmonium states to the $S$-wave charmonium states. There we
normalize $ \Gamma_{\chi_{c2}  \to \psi\gamma}=1$, the wide-hat
denotes the experiential values. }
\end{table}

\begin{figure}
 \centering
 \includegraphics[totalheight=6cm,width=8cm]{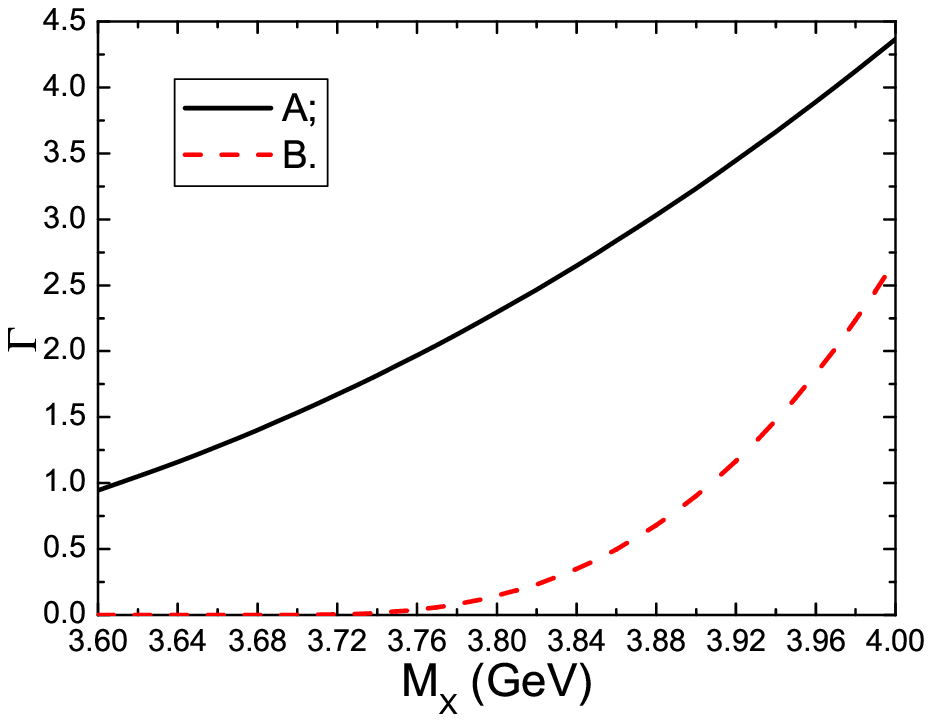}
  \caption{The radiative decay widths of the $X(3872)$  versus  its mass, the $A$ and $B$ denote the processes
 $X(3872) \to J/\psi \gamma$ and $X(3872) \to \psi' \gamma$, respectively; the units of the decay widths are $10^{-2}\delta^2(2,1)$ and $10^{-3}\delta^2(2,2)$, respectively, $\delta(2,2)\gg\delta(2,1)$. }
\end{figure}

\section{Conclusion}
In this article, we study the radiative decays  among the charmonium
states with the heavy quark effective theory, and make predictions
for ratios among the radiative decay widths of an special multiplet
to another multiplet, where the unknown couple constants
$\delta(m,n)$ are canceled out with each other. The predictions can
be confronted with the experimental data in the future at the
BESIII, KEK-B, RHIC, $\rm{\bar{P}ANDA}$ and LHCb,  and put
additional constraints  in identifying  the $X$, $Y$, $Z$
charmonium-like mesons.
\section*{Acknowledgment}
This  work is supported by National Natural Science Foundation of
China, Grant Number 11075053,   and the
Fundamental Research Funds for the Central Universities.

\end{document}